\documentclass[paper]{ieice}
\usepackage[dvipdfmx]{graphicx,xcolor}
\usepackage[fleqn]{amsmath}
\usepackage{newtxtext}
\usepackage[varg]{newtxmath}

\DeclareGraphicsExtensions{.eps}

\makeatletter
\newif\ifhbonecolumn
\@ifclasswith{ieice}{brief}{\hbonecolumntrue}{\hbonecolumnfalse}
\makeatother

\field{D}
\title[Diversity-Robust Acoustic Feature Signatures Based on Multiscale Fractal Dimension]{Diversity-Robust Acoustic Feature Signatures \\ Based on Multiscale Fractal Dimension \\ for Similarity Search of Environmental Sounds}
\authorlist{%
 \authorentry[sunouchi@media.scu.ac.jp]{Motohiro Sunouchi}{m}{aff1}\MembershipNumber{2014667}
 \authorentry[yoshioka@ist.hokudai.ac.jp]{Masaharu Yoshioka}{m}{aff2}\MembershipNumber{2014667}
}
\affiliate[aff1]{The author is with the Design Department, Sapporo city university, Geijutsu-no-mori 1, Minami-ku, Sapporo, Hokkaido 005--0864, Japan}
\affiliate[aff2]{The author is with the Graduate School of Information Science and Technology, Hokkaido University, Kita 14, Nishi 9, Kita--ku, Sapporo, Hokkaido, 060--0814, Japan}
\begin{document}
\maketitle
\begin{summary}
This paper proposes new acoustic feature signatures based on the multiscale fractal dimension (MFD), which are robust against the diversity of environmental sounds, for the content-based similarity search. The diversity of sound sources and acoustic compositions is a typical feature of environmental sounds.
Several acoustic features have been proposed for environmental sounds. Among them is the widely-used Mel-Frequency Cepstral Coefficients (MFCCs), which describes frequency-domain features. However, in addition to these features in the frequency domain, environmental sounds have other important features in the time domain with various time scales.
In our previous paper, we proposed enhanced multiscale fractal dimension signature (EMFD) for environmental sounds. This paper extends EMFD by using the kernel density estimation method, which results in better performance of the similarity search tasks. Furthermore, it newly proposes another acoustic feature signature based on MFD, namely very-long-range multiscale fractal dimension signature (MFD-VL). The MFD-VL signature describes several features of the time-varying envelope for long periods of time. The MFD-VL signature has stability and robustness against background noise and small fluctuations in the parameters of sound sources, which are produced in field recordings. We discuss the effectiveness of these signatures in the similarity sound search by comparing with acoustic features proposed in the DCASE 2018 challenges. Due to the unique descriptiveness of our proposed signatures, we confirmed the signatures are effective when they are used with other acoustic features.
\end{summary}

%%%%%%%%%%%%%%%%%%%%%%%%%%%%%%%%%%%%%%%%

\section{Introduction}

Acoustic feature extraction is a basic audio signal processing issue. Acoustic features are important and necessary for various contexts and applications related to environmental sound recognition (ESR), such as large-scale content-based retrieval, auditory scene analysis, visualization, and event detection for surveillance. During the last decade, handy digital sound recorders have gained popularity, and at present, not only professional creators, but also amateurs have started recording environmental sounds and sharing them on web services such as Freesound~\cite{Akkermans2011,freesound} and SoundCloud~\cite{soundcloud}. These sound recordings are not only appreciated as music works, but also sampled for creating sound effects, new music works, and live performances in music genres such as ambient, drone, and electronic~\cite{Lopez1998}~\cite{Michael2011}. These sound recordings are also utilized for research to analyze and understand the variety of sound environments that we live in~\cite{Park2014}.

\subsection{Applications Using Acoustic Features for Environmental Sounds}
In recent years, the research on ESR for understanding a scene and its context has received considerable attention~\cite{Chachada2013}. The workshop challenges on \textit{Detection and Classification of Acoustic Scenes and Events} (DCASE) have demonstrated performance evaluations of systems for the detection and classification of sound events~\cite{Stowell2015}. Based on the best result from Task 1B of DCASE2020, Koutini \textit{et al.}\ evaluated their Receptive Field (RF) regularized CNN model with some parameter reduction methods.~\cite{Koutini2020}.

Classification is a basic application that uses acoustic features. In 2003, Cowling and Sitte~\cite{Cowling2003} presented a comprehensive comparative study of classification techniques that use various acoustic features for environmental sounds. They reported that the test patterns using each of the Mel-Frequency Cepstral Coefficients (MFCCs) and the continuous wavelet transform achieved the best recognition performance. In 2009 and 2012, Chu \textit{et al.}\ \cite{Chu2009} and Mogi \textit{et al.}\ \cite{Mogi2012} reported that recognition systems that use the Matching-Pursuit-based acoustic feature as a time-domain feature shows better classification performance than systems that use the popular MFCCs only as a frequency-domain feature. In 2013, Bauge \textit{et al.}\ \cite{Bauge2013} proposed a new acoustic feature for environmental sounds based on the scattering transform. This feature is robust against frequency transposition.

Content-based retrieval is another basic application that uses acoustic features. Web-based sound archives such as Freesound and SoundCloud are becoming popular and the amount of their sound content is increasing. The online users who utilize these sound archives can share and browse sound content by means of content-based retrieval. In 2008, Xue \textit{et al.}\ \cite{Xue2008} proposed a similarity search system, which employs a cluster-based indexing approach for environmental sounds. In 2010, Roma \textit{et al.}\ \cite{Roma2010} proposed a method for the retrieval of environmental sounds using the general sound-events taxonomy defined based on the principles of ecological acoustics. Chechik \textit{et al.}\ \cite{Chechik2008} compared the scalability of several classification methods using MFCCs for a large-scale content-based sound retrieval. In 2013, Sunouchi and Tanaka~\cite{Sunouchi2013} proposed a new acoustic feature signature, namely, the enhanced multiscale fractal dimension signature (EMFD) and demonstrated the effectiveness of EMFD for content-based similarity search of environmental sounds.

In recent years, the workshop challenges on DCASE have focused on improving machine learning methods for ESR and produced high-performance results for their tasks. Acoustic features are still essential as input data for the machine learning methods for ESR. Hence finding new acoustic features that can properly describe the features of environmental sounds is fundamental in improving the performance of these ESR applications. In addition, by studying how acoustic features can describe the features of environmental sounds and affect the performance of ESR tasks, we can develop an understanding of how we are listening to environmental sounds.

\subsection{Acoustic Feature Extraction for ESR}\label{subsec_acousticfeatureextraction_for_esr}
The environmental sounds outside a recording studio  are produced by action and movement. We can identify things by listening to their acoustic properties, which are the results of the sound production process. However, environmental sound signals of the same type cannot be physically identical to each other due to the difference in their production processes. Furthermore, the different sound signals generated by simultaneous events are mixed with each other, which makes the properties of each sound source obscured~\cite{Handel1995}.

Various acoustic features have been proposed for content-based audio retrieval. The feature selection is an important process for ESR~\cite{Mitrovic2009}, ~\cite{Mitrovic2010}. Cepstral features that include MFCCs and their first and second derivatives (MFCCs$\Delta$ and MFCCs$\Delta\Delta$) are widely used as frequency-domain acoustic features. MP-based acoustic feature has been proposed as one of the useful time-domain features for ESR~\cite{Chu2009},\cite{Mogi2012}~\cite{Mallat1993}.

Recent researches have focused on the evaluation of time-domain features of environmental sounds. For ESR, we need acoustic features that describe the non-stationary characteristics of target sounds as a time-domain feature and are robust against the diversity of environmental sounds~\cite{Chachada2013}. We have recognized there may be three main causes of the diversity of environmental sounds.

\begin{itemize}

\item[D1)]Small fluctuations of sound source parameters, such as carrier signal frequency, due to the individuality of the sound source.
\item[D2)]Background noises that the person who recorded the target sound did not expect to record.
\item[D3)]Mixed composition of different types of sound sources.
\end{itemize}

For the third cause D3, it is necessary to apply, for example, independent component analysis or non-negative matrix factorization to the sound signal before the feature extraction process~\cite{Mogi2012},\cite{Innami2012}. In this study, we focus on the extraction of new acoustic feature signatures that are robust against the diversities caused by D1 and D2.

\subsection{Problems of EMFD Signature and Their Solutions}\label{subsec_problems}
In our previous work~\cite{Sunouchi2013}, we proposed an EMFD signature that can describe both the frequency-domain features and time-domain features of target sounds. The EMFD signature is a feature vector, which consists of the time-varying multiscale fractal dimension (MFD) values. We demonstrated that EMFD improves the performance of similarity search by supplementing MFCCs. Unfortunately, it is found that EMFD includes error values that depend on the number of analysis windows and the histogram's bin size used for computing its histogram. Furthermore, EMFD seems to be oversensitive while discriminating the features of environmental sounds, and may lack robustness against the diversity of environmental sounds.

In this study, we extend the EMFD signature by improving the process of computing its histogram using the kernel density estimation method. By optimizing the bandwidth parameter used for kernel density estimation, the histogram of the enhanced multiscale fractal dimension using kernel density estimation signature (EMFD-KDE) becomes sufficiently smooth and robust against the diversity of environmental sounds as an acoustic feature signature. In Sect.~\ref{sec_EMFD}, we present the basic theory and characteristics of EMFD. In Sect.~\ref{sec_EMFD-KDE}, we propose a method to compute the EMFD-KDE signature. In Sect.~\ref{sec_exp_eval} and \ref{sec_expevaldcase2018}, we demonstrate that EMFD-KDE improves the performance of the similarity search system.

Furthermore, we enhance the idea of EMFD and propose a new acoustic feature signature, namely very-long-range multiscale fractal dimension signature (MFD-VL). The environmental sounds have important acoustic features over a long time period. However, EMFD cannot describe the time-domain feature for time periods longer than 10 ms. In Sect.~\ref{sec_MFD-VL}, we propose a method to compute the MFD-VL signature. In addition, we demonstrate that MFD-VL can describe the features of the time varying envelope for long periods of time, and that it has the robustness against the diversity causes D1 and D2.

In Sect.~\ref{sec_conclusion}, we conclude that the proposed feature signatures of EMFD-KDE and MFD-VL solve the problems of EMFD and are effective when they are used with other acoustic features, including MFCCs and acoustic features proposed in the DCASE 2018 challenges.

\section{Basic Theory of Enhanced Multiscale Fractal Dimension Signature}\label{sec_EMFD}
Mandelbrot, who advocated a concept of fractal in 1975 for the first time, demonstrated that some structures in nature could be modeled well by the theory of fractals~\cite{Mandelbrot1982}. One of the most important characteristics of fractals is that they have self-similarity properties at multiple scales. In the field of acoustics, Voss and Clarke analyzed the power spectrum of fluctuating physical variables including frequency, loudness and pitch in music and speech\cite{VOSS1975}. They obtained the $1/f^{\gamma}~(0.5 \lesssim \gamma \lesssim 1.5)$ aspects in the power spectrum of each variable against the frequency of a signal passed through a low-pass filter having a range 0\,Hz -- 1\,Hz. Hsu~\cite{Hsu1990a} compared the fractal geometry of classical music works, and found that there is a relation, defined by the theory of fractals, between the interval of successive notes and their frequency of occurrence.

\subsection{Multiscale Fractal Dimension}
A fractal dimension is an index value that can describe the characteristics of a fractal by quantifying their complexity as a ratio of the change in detail to the change in scale. Acoustic features based on the fractal dimension have been proposed and utilized for various practical applications in the fields such as acoustics, music analysis, image analysis, physics, physiology, and neuroscience. Maragos \textit{et al.}\ \cite{Maragos1991}\cite{Maragos1999} proposed the short-time fractal dimension of speech signals as an acoustic feature and used it for speech segmentation and sound classification. Zlatintsi and Maragos~\cite{Zlatintsi2011}\cite{Zlatintsi2013} proposed a multiscale fractal dimension (MFD) profile as a short-time descriptor and found that this descriptor can discriminate several aspects among different musical instruments.

\subsection{Steps to Compute the EMFD Signature}\label{subsec_EMFD}
In our previous work~\cite{Sunouchi2013}, we developed EMFD as a feature signature of environmental sounds for a similarity search system. The EMFD is computed as follows.

\subsubsection{Preprocessing Target Sounds}
The maximum amplitude of each target sound that is to be analyzed must be first normalized to -0.1\,db. They are converted to the standard format with the following specifications: sampling rate of 44.1\,kHz and bit depth of 16\,bits.

\subsubsection{Computing the Area of Minkowski Sausage}
\begin{figure}[!t]
\centering
\includegraphics[width=2.8in]{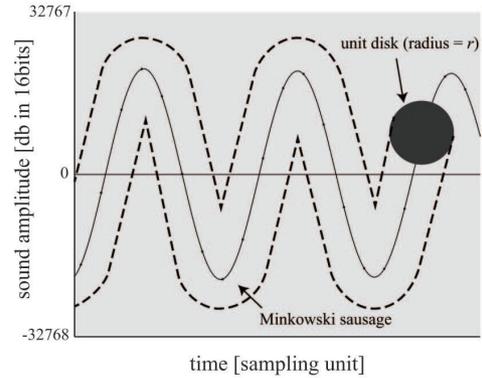}
\caption{A sound waveform and a Minkowski sausage}
\label{fig_Minkowski}
\end{figure}

\begin{figure}[!t]
\centering
\includegraphics[width=2.8in]{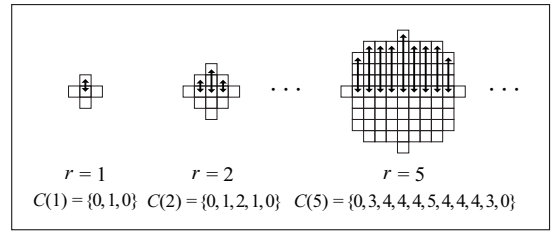}
\caption{Mesh-Approximation of a unit disk}
\label{fig_unitdisk}
\end{figure}

The fractal dimension of a sound signal can be computed based on the Minkowski-Bouligand dimension. A covering area can be drawn by moving a unit disk of radius $r$ along the curve of the waveform. This covering area is called a Minkowski Sausage. The center of the unit disk should be at any position on the curve of waveform and the width of Minkowski Sausage becomes $2r$. Figure~\ref{fig_Minkowski} shows the Minkowski Sausage obtained by moving the unit disk along the waveform. To compute the area of the Minkowski Sausage of a discrete sound signal, the unit disk vector $C(r)$ is defined as Eq.~(\ref{eq_unitdiskvector}), where $r$ denotes the radius of the unit disk and $i$ denotes the discrete position on the horizon. Figure~\ref{fig_unitdisk} shows how the model of the unit disk is built. The vertical distance from the center to the top of unit disk at each horizontal position is denoted by the unit disk vector $C(r)$. Let $n$ be the sampling position, $r$ the radius of the unit disk, $p$ the discrete position of the unit disk, and $\mathrm{sig}(x)$ the amplitude of sound signal at each sampling position $x$. The area of the Minkowski Sausage $area(n, r)$ at each sampling position $n$ is computed as Eq.~(\ref{eq_minkowskiarean}).

\begin{equation}
C(r) = \left\{ \mathrm{floor}\left( \sqrt{2ri-i^2} \right) \Bigm| 0 \le i \le 2r, i \in \mathbb{Z} \right\}
\label{eq_unitdiskvector}
\end{equation}

\ifhbonecolumn
\begin{multline}
\mathrm{area}(n, r) = \max\limits_{\substack{0 \le p \le 2r \\ p \in \mathbb{Z}}}\left( \mathrm{sig}\left( n - r + p \right) + \mathrm{floor}\left( \sqrt{2rp-p^2} \right) \right) \\
-\min\limits_{\substack{0 \le p \le 2r \\ p \in \mathbb{Z}}}\left( \mathrm{sig}\left( n - r + p \right) - \mathrm{floor}\left( \sqrt{2rp-p^2} \right) \right)
\label{eq_minkowskiarean}
\end{multline}

\else
\begin{multline}
\mathrm{area}(n, r)\!=\!\max\limits_{\substack{0 \le p \le 2r \\ p \in \mathbb{Z}}}\!\left(\!\mathrm{sig}\left(n\!-\!r\!+\!p\right)\!+\!\mathrm{floor}\!\left(\!\sqrt{2rp\!-\!p^2}\!\right)\!\right) \\
\!-\!\min\limits_{\substack{0 \le p \le 2r \\ p \in \mathbb{Z}}}\!\left(\!\mathrm{sig}\left(n\!-\!r\!+\!p\right)\!-\!\mathrm{floor}\!\left(\!\sqrt{2rp\!-\!p^2}\!\right)\!\right)
\label{eq_minkowskiarean}
\end{multline}

\fi

\subsubsection{Definition of Multiscale Fractal Dimension}
The MFD values are computed for each analysis window whose period is 50ms. Let $A(r)$ be the area of the Minkowski Sausage drawn by the unit disk of radius $r$ in each analysis window. The MFD of each analysis window is defined by Eq.~(\ref{eq_mfd}). The minimum radius (r=1) corresponds to the sampling period of the signal (1/44.1~ms) and the range of $r$ from 1 to 132 corresponds to the range of the time scales from 1/44.1 to 3~ms.

\mathindent=0mm
\begin{equation}
\mathop{MFD}\!=\!\left\{2\!-\!\frac{\log\left( A(r\!+\!1)/A(r) \right)}{\log\left( (r\!+\!1)/r \right)} \Biggm| 1 \!\le\! r \!\le\! 132,~r \!\in\! \mathbb{Z} \right\}
\label{eq_mfd}
\end{equation}
\mathindent=7mm

\ifhbonecolumn

\begin{figure*}[!t]
\normalsize

\begin{equation}
\mathop{MFD_{enhanced}(x)} = 2-\frac{\log\left( A(r(x+1)) / A(r(x)) \right)}{\log\left( r(x+1)/r(x) \right)} \mathrm{, where} ~r(x) = \mathrm{round}(1.4^x)
\label{eq_emfd}
\end{equation}

\begin{equation}
\mathop{AW(sound,~period)} = \left\{ 0, period, ... , \mathrm{floor}\left(\frac{\mbox{the length of}~sound}{period} - 1\right) \times period \right\}
\label{eq_analysiswindow}
\end{equation}

\begin{multline}
\mathop{FAW(dbin,~rbin)} = \bigl\{~t~\bigm|~t\in{AW(sound,~50)},
1+(dbin-1)/32 \le MFD_{enhanced}(rbin)~\mbox{of analysis window}~t < 1+dbin/32 \bigr\}
\label{eq_EMFD_analysiswindownum}
\end{multline}

\begin{multline}
\mathop{EMFD(sound)} =
\left\{\frac{\mathrm{card}\left(FAW(dbin, rbin)\right)}{\mathrm{card}\left(AW(sound,~50)\right)}\Biggm| \right. \\
1 \le dbin \le 32,~dbin\in\mathbb{Z},~1 \le rbin \le 16,~rbin\in\mathbb{Z} \Biggr\}
\label{eq_EMFDdefinition}
\end{multline}

\hrulefill

\normalsize
\end{figure*}

\else %%%%%%%

\begin{figure*}[!t]
\normalsize

\begin{equation}
\mathop{MFD_{enhanced}(x)} = 2-\frac{\log\left( A(r(x+1)) / A(r(x)) \right)}{\log\left( r(x+1)/r(x) \right)} \mathrm{, where} ~r(x) = \mathrm{round}(1.4^x)
\label{eq_emfd}
\end{equation}

\begin{equation}
\mathop{AW(sound,~period)} = \left\{ 0, period, ... , \mathrm{floor}\left(\frac{\mbox{the length of}~sound}{period} - 1\right) \times period \right\}
\label{eq_analysiswindow}
\end{equation}

\begin{multline}
\mathop{FAW(dbin,~rbin)} = \bigl\{~t~\bigm|~t\in{AW(sound,~50)},\\
1+(dbin-1)/32 \le MFD_{enhanced}(rbin)~\mbox{of analysis window}~t < 1+dbin/32 \bigr\}
\label{eq_EMFD_analysiswindownum}
\end{multline}

\begin{equation}
\mathop{EMFD(sound)} =
\Biggl\{\frac{\mathrm{card}\left(FAW(dbin, rbin)\right)}{\mathrm{card}\left(AW(sound,~50)\right)}\Biggm| 1 \le dbin \le 32,~dbin\in\mathbb{Z},~1 \le rbin \le 16,~rbin\in\mathbb{Z} \Biggr\}
\label{eq_EMFDdefinition}
\end{equation}
\hrulefill

\normalsize
\end{figure*}

\fi

\subsubsection{Definition of the EMFD Signature}
\begin{figure}[!t]
\centering
\includegraphics[width=2.4in]{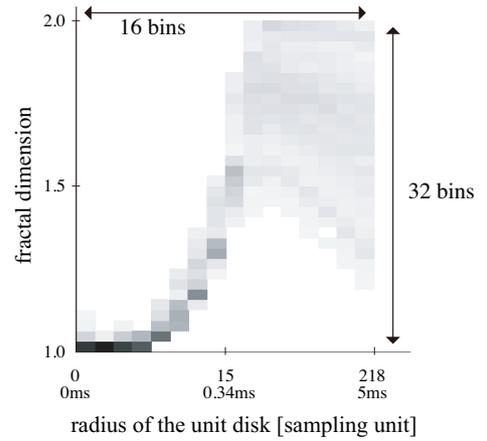}
\caption{Visualization of the EMFD signature of a cuckoo sound $S_{cuckoo}$\label{fig_emfdbincuckoo}}
\end{figure}

In our previous work~\cite{Sunouchi2013}, we found that MFD has informative values for unit disks larger than the disk with a radius of 3~ms $(r=132)$. The maximum radius of the unit disk was extended to 218, which corresponds to 5~ms (1/10 of the period of analysis window), and the discrete values of the unit disk were modified to have exponential values. The enhanced MFD value at the \mbox{$x$-th} discrete value of the unit disk is defined as Eq.~(\ref{eq_emfd}). The enhanced MFD values are computed for each analysis window. The EMFD signature is then defined as the two-dimensional histogram $(16\times32)$ of the time-varying enhanced MFD. Let $period$ be the period (ms) of the analysis window and $sound$ be the target sound. The set of analysis windows of the target sound is defined as Eq.~(\ref{eq_analysiswindow}). Let $rbin$ be the bins that correspond to a series of 16 numbers used to define the different radius of the unit disk for computing the enhanced MFD, and $dbin$ be the bins that correspond to a series of the 32 small intervals into which the range of fractal dimension is divided. The set of analysis windows whose enhanced MFD values fall into the bin $(dbin, rbin)$ is defined by Eq.~(\ref{eq_EMFD_analysiswindownum}). The set of values in each bin of the EMFD histogram is defined by Eq.~(\ref{eq_EMFDdefinition}). Figure~\ref{fig_emfdbincuckoo} shows the histogram that visualizes the EMFD signature of a cuckoo sound $S_{cuckoo}$. The length of $S_{cuckoo}$ is 21.08s.

\subsection{Known Characteristics of the EMFD signature}
Zlatintsi and Maragos~\cite{Zlatintsi2013} concluded that the MFD profiles are useful for quantifying the multiscale complexity and fragmentation of the different states of the instrument sound waveforms. In our previous work~\cite{Sunouchi2013}, we confirmed that EMFD describes the frequency-domain features and several other effective features of environmental sounds that MFCCs cannot describe. Furthermore, we confirmed that the EMFD signature has robustness against changes in volume levels and phase shifting of sound signals in the analysis window.

%%%
%%% deleted
%%%

\subsection{Problems of the EMFD signature}
The EMFD includes error values that depend on the number of analysis windows and the histogram's bin size, which is defined for computing its histogram, as shown in Eq.~(\ref{eq_EMFDdefinition}). In Sect.~\ref{sec_EMFD-KDE}, we extend the existing EMFD by employing a kernel density estimation method to solve this problem.

Another problem of EMFD is that it cannot describe time-domain features for a time period longer than 10~ms, although environmental sounds have important acoustic features over a long time period. To solve this problem, we introduce MFD-VL signature in Sect.~\ref{sec_MFD-VL} as a newly developed time-domain acoustic feature.

\section{Extending EMFD Employing a Kernel Density Estimation Method}\label{sec_EMFD-KDE}
As mentioned in Subsect.~\ref{subsec_EMFD}, EMFD is computed as the two-dimensional histogram of time-varying enhanced MFD values. Let $NFAW(bin)$ be the number of analysis windows whose enhanced MFD values fall into the bin, and $NAW(sound)$ be the total number of analysis windows of the target sound $sound$, as defined in Eq.~(\ref{eq_numberofaw}). The EMFD value of each bin $EMFD(sound,~bin)$ is computed as Eq.~(\ref{eq_EMFDeachbin}). This method has the following two problems.

\begin{equation}
\mathop{NAW(sound)} = \mathrm{card}\bigl(AW(sound,~50)\bigr)
\label{eq_numberofaw}
\end{equation}

\begin{equation}
\mathop{EMFD(sound,~bin)} = \frac{NFAW(bin)}{NAW(sound)}
\label{eq_EMFDeachbin}
\end{equation}

The first problem is that the value of each bin necessarily includes an error, because the value can be one of the discrete values given by the density of analysis windows. In particular, a lower number of analysis windows for the target sound increases the errors. Ideally, the EMFD histogram should be a continuous probability distribution of the time-varying enhanced MFD values, regardless of the number of analysis windows.

The second problem is that EMFD computed by the existing method is oversensitive to discriminate the features of environmental sounds. The tones and frequencies of each environmental sound may often vary, depending on the recording conditions and individual characteristics of the sources that generate this sound, even if person try to record the same type of environmental sounds in the same way. Therefore, a feature signature of environmental sounds should have robustness against the diversity of environmental sounds caused by D1 defined in Subsect.~\ref{subsec_acousticfeatureextraction_for_esr}.

To solve these problems, we introduce the kernel density estimation method to compute the EMFD histogram.

\subsection{Definition of the EMFD-KDE Signature}
The kernel density estimation method is employed to compute the probability distribution of the enhanced MFD values at each radius of the unit disk. The values of each bin of EMFD-KDE are defined as shown in Eq.~(\ref{eq_emfd}), Eq.~(\ref{eq_numberofaw}), Eq.~(\ref{eq_kernel}), Eq.~(\ref{eq_kde}), and Eq.~(\ref{eq_EMFDKDEdefinition}), where $K(\cdot)$ is the kernel function which is a Gaussian function, and $h$ in Eq.~(\ref{eq_kde}) is the smoothing parameter called bandwidth.

\begin{equation}
\mathop{K(x)} = \frac{1}{\sqrt{2\pi}}e^{-\frac{1}{2}x^2}
\label{eq_kernel}
\end{equation}

\ifhbonecolumn
\begin{equation}
\mathop{f_{emfd\mathchar`-kde}(dbin_{val}, rbin)} = \frac{1}{NAW~h} \times \sum_{}^{NAW}K\left(\frac{dbin_{val}-MFD_{enhanced}(rbin)}{h}\right)
\label{eq_kde}
\end{equation}

\begin{multline}
EMFD\mathchar`-KDE = \Biggl\{f_{emfd\mathchar`-kde}\left(1 + \frac{dbin - 0.5}{32}, rbin\right) \\
\Biggm| 1 \le dbin \le 32,~dbin\in\mathbb{Z},~1 \le rbin \le 16,~rbin\in\mathbb{Z}~\Biggr\}
\label{eq_EMFDKDEdefinition}
\end{multline}

\else
\begin{multline}
\mathop{f_{emfd\mbox{\scriptsize -}kde}(dbin_{val}, rbin)} = \frac{1}{NAW~h} \\
\times \sum_{}^{NAW}K\left(\frac{dbin_{val}-MFD_{enhanced}(rbin)}{h}\right)
\label{eq_kde}
\end{multline}

\mathindent=0mm
\begin{multline}
EMFD\mathchar`-KDE\!=\!\Biggl\{f_{emfd\mbox{\scriptsize -}kde}\left(1\!+\!\frac{dbin\!-\!0.5}{32}, rbin\right) \\
\Biggm| 1 \!\le\! dbin \!\le\! 32,~dbin \!\in\! \mathbb{Z}, 1\!\le\!rbin\!\le\!16,~rbin\!\in\!\mathbb{Z}~\Biggr\}
\label{eq_EMFDKDEdefinition}
\end{multline}
\mathindent=7mm

\fi

\subsection{Optimization of the Bandwidth for Kernel Density Estimation}\label{subsec_opt_bandwidth}
The bandwidth $h$ is a smoothing parameter, which is usually determined by the trade-off between the number of data samples and their standard deviation. Let $n$ be the number of data samples and $\sigma$ be the standard deviation of the data samples. The bandwidth $h$ of a Gaussian kernel density estimator is given by the normal reference rule defined by Eq.~(\ref{eq_bwscott}). The normal reference rule is most commonly used to determine the bandwidth \cite{Scott1992}.

\begin{equation}
\mathop{h} = \left(\frac{4\sigma^5}{3n}\right)^{\frac{1}{5}} \approx 1.06 \sigma n^{-\frac{1}{5}}
\label{eq_bwscott}
\end{equation}

We define the bandwidth $h_{rbin}(\alpha)$, which is optimized for each radius of the unit disk, as Eq.~(\ref{eq_bwEMFDavg}), Eq.~(\ref{eq_bwEMFDsigma}), and Eq.~(\ref{eq_bwEMFD}), where $avg$ is the arithmetic mean of the enhanced MFD values of each analysis window at $rbin$, and $\sigma_{rbin}$ is the standard deviation of the enhanced MFD values at $rbin$. The smoothing parameter $\alpha$ in Eq.~(\ref{eq_bwEMFD}) is a constant. Through the experiments with different values of $\alpha$, the constant value is determined so that the best result for the target task is obtained.

%%%%%% Through the experiments with different values of $\alpha$, we found that the best result for the similarity search is obtained for $\alpha = 32$. In this study, $h_{rbin}(32)$ is used as the bandwidth for each radius of the unit disk to compute the EMFD-KDE signature.

\begin{equation}
avg = \frac{1}{NAW}\sum_{}^{NAW}{MFD_{enhanced}(rbin)}
\label{eq_bwEMFDavg}
\end{equation}

\mathindent=0mm
\begin{equation}
\sigma_{rbin}\!=\!\sqrt{\frac{1}{NAW}\!\sum_{}^{NAW}\!\!\left(\!MFD_{enhanced}(rbin)\!-\!avg\right)^2}
\label{eq_bwEMFDsigma}
\end{equation}
\mathindent=7mm

\begin{equation}
\mathop{h_{rbin}(\alpha)} = 1.06 \sigma_{rbin} NAW^{-\frac{1}{5}} \alpha
\label{eq_bwEMFD}
\end{equation}

\begin{figure}[!t]
\centering
\includegraphics[width=2.8in]{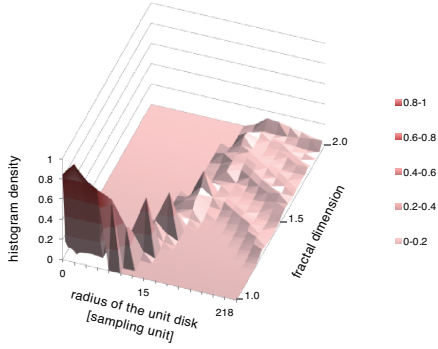}
\caption{The 3D histogram image that visualizes the existing EMFD signature of the cuckoo sound $S_{cuckoo}$}
\label{fig_emfdbin3d}
\end{figure}

\begin{figure}[!t]
\centering
\includegraphics[width=2.8in]{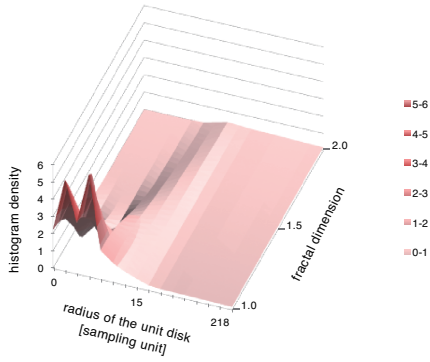}
\caption{The 3D histogram image that visualizes the EMFD-KDE signature of the cuckoo sound $S_{cuckoo}$. The bandwidth is $h_{rbin}(32)$.}
\label{fig_emfdkde3d}
\end{figure}

Figure~\ref{fig_emfdbin3d} shows the 3D histogram visualizing the EMFD signature of the cuckoo sound $S_{cuckoo}$. Figure~\ref{fig_emfdkde3d} shows the 3D histogram visualizing the EMFD-KDE signature of the same cuckoo sound $S_{cuckoo}$. The 3D histogram of the EMFD-KDE signature is much smoother than that of the EMFD signature. At each radius of the unit disk, the larger standard deviation of the enhanced MFD values $\sigma_{rbin}$ results in the smoother histogram.

\section{Very Long Range Multiscale Fractal Dimension Signature}\label{sec_MFD-VL}
The environmental sounds have important acoustic features with varying time periods. However, EMFD cannot describe the time-domain features for time periods longer than 10~ms. To solve this problem, we propose a new acoustic feature signature for environmental sounds, based on the multiscale fractal dimension. This feature signature is called very-long-range multiscale fractal dimension signature (MFD-VL). The basic idea of MFD-VL is to extend the size range of the unit figure to consider the larger ones, which are used to compute the area of the Minkowski Sausage. In this section, we define the method to compute the MFD-VL signature and demonstrate its characteristics.

\subsection{Definition of the MFD-VL Signature}
\begin{figure}[!t]
\centering
\includegraphics[width=2.8in]{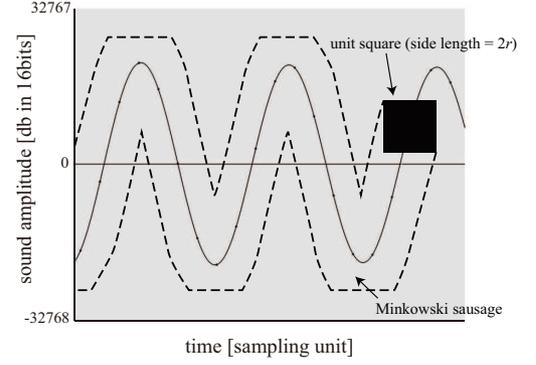}
\caption{A sound waveform and a Minkowski Sausage obtained by moving the unit square}
\label{fig_MinkowskiSquare}
\end{figure}

The multiscale fractal dimension values of MFD-VL are computed for an entire target sound, and not for each fixed-length analysis window of the target sound. A unit square, instead of a unit disk, is used to compute the area of the Minkowski Sausage for MFD-VL. Figure~\ref{fig_MinkowskiSquare} shows the Minkowski Sausage obtained by moving the unit square along the waveform. The method using the unit square is much faster than the one using the unit disk. Let $n$ denote the sampling position, $r$ the half side-length of the unit square, $p$ the discrete position of the unit square, and $sig(x)$ the amplitude value of the sound signal at each sampling position $x$. The area of the Minkowski Sausage $area_{sq}(n, r)$ at each sampling position $n$ is computed as Eq.~(\ref{eq_minkowskiareansq}).

\ifhbonecolumn
\begin{equation}
area_{sq}(n, r) = \max\limits_{\substack{0 \le p \le 2r \\ p \in \mathbb{Z}}}\left( \mathrm{sig}\left( n - r + p \right) + r \right) - \min\limits_{\substack{0 \le p \le 2r \\ p \in \mathbb{Z}}}\left( \mathrm{sig}\left( n - r + p \right) - r \right)
\label{eq_minkowskiareansq}
\end{equation}

\else
\begin{multline}
area_{sq}(n, r) = \max\limits_{\substack{0 \le p \le 2r \\ p \in \mathbb{Z}}}\left( \mathrm{sig}\left( n - r + p \right) + r \right) \\
-\min\limits_{\substack{0 \le p \le 2r \\ p \in \mathbb{Z}}}\left( \mathrm{sig}\left( n - r + p \right) - r \right)
\label{eq_minkowskiareansq}
\end{multline}

\fi

The half side-length $r$ of the unit square for each scale was defined as Eq.~(\ref{eq_MFDVLr}), where $sf$ is the sampling frequency of the target sound. In this study, $sf$ is 44100~Hz. Let $A_{sq}(r)$ denote the area of the Minkowski Sausage obtained for an entire target sound by moving the unit square whose side length is $2r$. The MFD-VL signature is defined as Eq.~(\ref{eq_mfdvl}). The MFD-VL signature is a feature vector that contains 10 elements.

\begin{equation}
r(x)=\mathrm{round}\left(sf \times 2^{-{\frac{x+2}{2}}}\right)
\label{eq_MFDVLr}
\end{equation}

\ifhbonecolumn
\begin{equation}
MFD\mathchar`-VL = \biggl\{ 2-\frac{\log\left( A_{sq}(r(x)) / A_{sq}(r(x+1)) \right)}{\log\left( r(x)/r(x+1) \right)} \biggm| 0 \le x \le 9,~x\in\mathbb{Z} \biggr\}
\label{eq_mfdvl}
\end{equation}

\else
\begin{multline}
MFD\mathchar`-VL\!=\!\biggl\{ 2 \!-\! \frac{\log\left( A_{sq}(r(x)) / A_{sq}(r(x\!+\!1)) \right)}{\log\left( r(x)/r(x\!+\!1) \right)} \\
\biggm| 0 \!\le\! x \!\le\! 9,~x \!\in\! \mathbb{Z} \biggr\}
\label{eq_mfdvl}
\end{multline}

\fi

\subsection{Basic Characteristics of the MFD-VL Signature}
We found several basic characteristics of MFD-VL through the experiments using test sounds.

\subsubsection{MFD-VL's Descriptiveness of the Beats of Single Sine Waves}\label{subsubsec_MFD-VLdescbeatssw}
The MFD-VL signature is expected to describe the acoustic features over very long time-periods. We found that MFD-VL can discriminate frequencies of amplitude envelopes between 22.6~Hz and 1~Hz. The range of the wavelength of the amplitude envelopes corresponds to the range of the side length of the unit square between 0.044~s and 1~s. Let $f_{beat}$ denote the frequency of the beats and $f_{content}$ denote the frequency of single sine waves inside the amplitude envelopes. The set of test sounds $SS_{t1}$ is defined as Eq.~(\ref{eq_addpinknoise}), Eq.~(\ref{eq_unaritest}), and Eq.~(\ref{eq_unaritestset}). Each test sound is filtered by the pink noise filter function $f_{pn}$ Eq.~(\ref{eq_addpinknoise}). A sound that is artificially synthesized using pure tones usually has distinct or sparse spectra. This kind of sounds may cause numerical instabilities while calculating their acoustic features. To solve this problem, the pink noise filter function $f_{pn}$ is used to add a background pink noise, which is defined as Eq.~(\ref{eq_addpinknoise}), where $sig$ is an input signal and $Noise_{pink}$ is a background pink noise whose maximum amplitude is normalized to -0.1~db. The signal-to-noise ratio is 24 db. The pink noise, known as $1/f$ noise, is a signal whose power spectral density is inversely proportional to the signal frequency. The pink noise signal is known to widely exist in the natural world. The frequency components below 40 Hz contained in the pink noise are cut off by using a low cut filter before the amplitude normalization because the components with lower frequencies cannot be recorded nor played using common microphones and speakers.

\begin{equation}
\mathop{f_{pn}\left(sig\right)} = \frac{15}{16}sig + \frac{1}{16}Noise_{pink}
\label{eq_addpinknoise}
\end{equation}

\mathindent=0mm
\begin{equation}
s_{t1}(f_{beat}, f_{content})\!=\!f_{pn}\bigl(\cos(\pi{f_{beat}}~t)\sin(2\pi{f_{content}}~t)\bigr)
\label{eq_unaritest}
\end{equation}

\ifhbonecolumn
\begin{equation}
SS_{t1} = \bigl\{s_{t1}(f_{beat}, f_{content}) \bigm| f_{content} = 440, f_{beat}\in\{0.5, 1, 2, 4, 8, 16, 32\}\bigr\}
\label{eq_unaritestset}
\end{equation}

\else
\begin{equation}
\begin{split}
SS_{t1} = &\bigl\{s_{t1}(f_{beat}, f_{content}) \\
& \quad \bigm| f_{content} = 440, f_{beat}\in\{0.5, 1, 2, 4, 8, 16, 32\}\bigr\}
\label{eq_unaritestset}
\end{split}
\end{equation}

\fi
\mathindent=7mm

\begin{figure}[!t]
\centering
\includegraphics[width=3.3in]{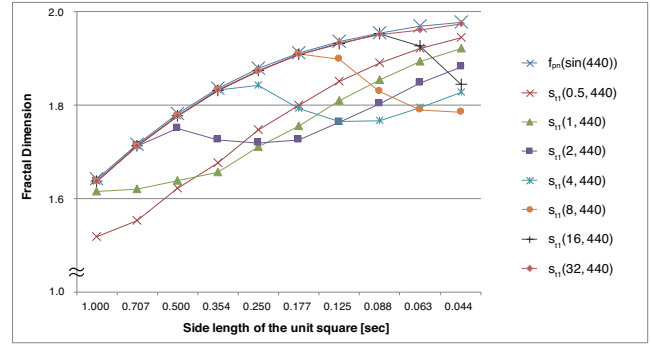}
\caption{Line charts of the MFD-VL signatures of single sine waves of frequency 440Hz, with and without a beat. The beat frequencies are 0.5Hz, 1Hz, 2Hz, 4Hz, 8Hz, 16Hz, and 32Hz.}
\label{fig_MFD-VL_UNARI}
\end{figure}

\begin{figure}[!t]
\centering
\includegraphics[width=2.5in]{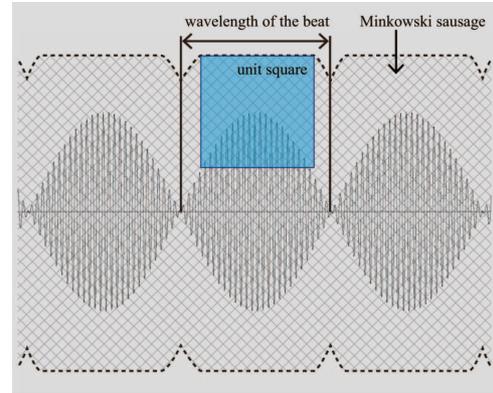}
\caption{Minkowski Sausage obtained by moving the unit square whose side length is less than the wavelength of the beat.}
\label{fig_MFD-VL_square}
\end{figure}

Figure~\ref{fig_MFD-VL_UNARI} shows the line charts of the MFD-VL values of single sine waves of frequency 440~Hz, which is filtered by Eq.~(\ref{eq_addpinknoise}), and those of the test sounds $SS_{t1}$. In Fig.~\ref{fig_MFD-VL_UNARI}, the frequencies of the beats are indicated by the troughs of the line chart, in which the side length of the unit square is less than the wavelengths of the beats. This characteristic can be understood morphologically as shown in Fig.~\ref{fig_MFD-VL_square}. When the side length of the unit square is more than the wavelength of the beat, the area of the Minkowski Sausage becomes almost the same as that of a single sine wave without beats. When the side length of the unit square is less than the wavelengths of the beats, the shorter side length of the unit square results in a smaller area of Minkowski Sausage.

\subsubsection{MFD-VL's Descriptiveness of Amplitude Envelope Shapes}
Here we analyze some other characteristics of the MFD-VL that are related to the descriptiveness of the amplitude envelope shapes.
Let $f_{pulse}$ be the frequency of the rectangular pulse waves and $w_{pulse}$ be the ratio of the rectangular pulse width to the wavelength of the rectangular pulse waves.
The rectangular pulse function $rect({f_{pulse}},~{w_{pulse}},~t)$ for generating the amplitude envelopes is defined as Eq.~(\ref{eq_MFD-VL-pulsetestrect}).
Let $f_{content}$ be the frequency of a single sine wave inside the amplitude envelopes generated by the rectangular pulse function.
The test sound $s_{t2}$ is defined as Eq.~(\ref{eq_MFD-VL-pulsetestrect}) and Eq.~(\ref{eq_MFD-VL-pulsetest}). The test sound is filtered by the pink noise filter function $f_{pn}$ Eq.~(\ref{eq_addpinknoise}).

\mathindent=0mm
\begin{equation}
rect({f_{pulse}},~{w_{pulse}},~t) \!=\! \begin{cases}
    1 \!&\! \left(t \bmod \frac{1}{f_{pulse}} <= \frac{w_{pulse}}{f_{pulse}}\right) \\
    0 \!&\! (otherwise)
  \end{cases}
\label{eq_MFD-VL-pulsetestrect}
\end{equation}

\ifhbonecolumn
\begin{equation}
s_{t2}(f_{pulse}, w_{pulse}, f_{content}) = f_{pn}\bigl(rect({f_{pulse}},~{w_{pulse}},~t)\sin(2\pi{f_{content}}~t)\bigr)
\label{eq_MFD-VL-pulsetest}
\end{equation}

\else
\begin{multline}
s_{t2}(f_{pulse}, w_{pulse}, f_{content}) \\
= f_{pn}\bigl(rect({f_{pulse}},~{w_{pulse}},~t)\sin(2\pi{f_{content}}~t)\bigr)
\label{eq_MFD-VL-pulsetest}
\end{multline}

\fi
\mathindent=7mm

\begin{figure}[!t]
\centering
\includegraphics[width=3.3in]{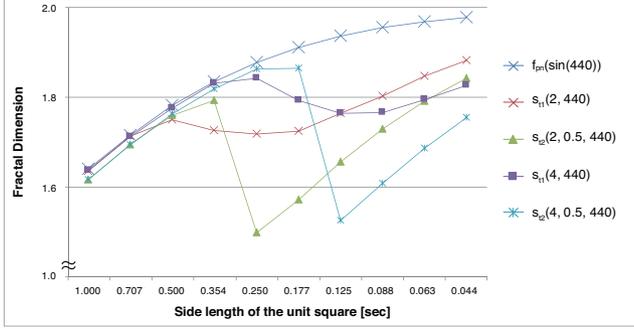}
\caption{Line charts of the MFD-VL signatures for single sine waves having a frequency of 440~Hz; those masked by the cosine function, and those masked by the rectangular pulse function.}
\label{fig_MFD-VL_SHAPE}
\end{figure}

We define the set of test sounds $SS_{t2}$ as Eq.~(\ref{eq_testset_sst2}). The line charts of MFD-VL of the single sine wave of frequency 440~Hz, which is filtered by Eq.~(\ref{eq_addpinknoise}), and $SS_{t2}$ are showed in Fig.~\ref{fig_MFD-VL_SHAPE}. Here, we compare the line charts of MFD-VL for $s_{t1}$ and those for $s_{t2}$. This comparison shows that the bottom of the line chart trough of $s_{t2}$ for $f_{pulse} = 2$ is deeper than that of $s_{t1}$ for $f_{beat} = 2$, and that the bottom of the line chart trough of $s_{t2}$ for $f_{pulse} = 4$ is also deeper than that of $s_{t1}$ for $f_{beat} = 4$.

\ifhbonecolumn
\begin{multline}
SS_{t2} = \left\{s_{t1}(f_{beat}, f_{content}) \bigm| f_{content} = 440 , f_{beat}\in\{2, 4\}\right\} \\
\cup \left\{s_{t2}(f_{pulse}, w_{pulse}, f_{content}) \bigm| f_{content} = 440 , f_{pulse}\in\{2, 4\} , w_{pulse} = 0.5\right\}
\label{eq_testset_sst2}
\end{multline}

\else
\mathindent=0mm
\begin{equation}
\begin{split}
SS_{t2} \!= &\bigl\{s_{t1}(f_{beat}, f_{content}) \!\bigm|\! f_{content} \!=\! 440 , f_{beat}\!\in\!\{2, 4\}\bigr\} \\
&\cup \bigl\{s_{t2}(f_{pulse}, w_{pulse}, f_{content}) \\
&\quad \bigm| f_{content} \!=\! 440 , f_{pulse}\!\in\!\{2, 4\} , w_{pulse} \!=\! 0.5 \bigr\}
\label{eq_testset_sst2}
\end{split}
\end{equation}
\mathindent=7mm

\fi

\begin{figure}[!t]
\centering
\includegraphics[width=3.3in]{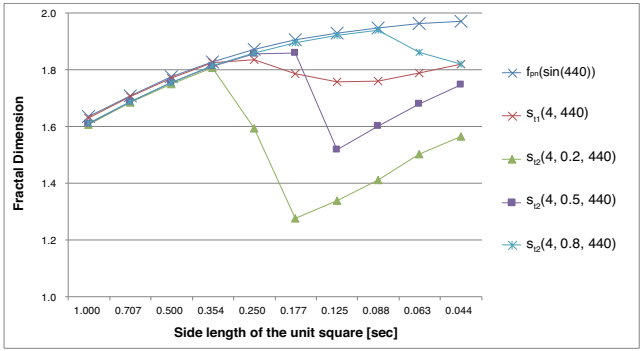}
\caption{Line charts of the MFD-VL signatures of single sine waves having a frequency of 440~Hz; those masked by the cosine function, and those masked by the rectangular pulse function having various widths.}
\label{fig_MFD-VL_PULSE-WIDTH}
\end{figure}

For another comparison, we define a set of test sounds $SS_{t3}$ as Eq.~(\ref{eq_pulsetestset}). The set of test sounds $SS_{t3}$ contains single sine waves having a frequency of 440~Hz masked by the rectangular pulse functions with various widths ${w_{pulse}}$. In Fig.~\ref{fig_MFD-VL_PULSE-WIDTH}, we compare the line charts of MFD-VL of single sine wave having a frequency of 440~Hz, $s_{t1} (f_{content}=440, f_{beat}=4)$, and $SS_{t3}$. This comparison shows that the narrower width of the amplitude envelopes made by the rectangular pulse function results in deeper troughs in the line chart.

\ifhbonecolumn
\begin{equation}
SS_{t3} = \bigl\{s_{t2}(f_{pulse}, w_{pulse}, f_{content}) \bigm| {f_{content}} = 440 , {f_{pulse}} = 4 , {w_{pulse}}\in\{0.2, 0.5, 0.8\}\bigr\}
\label{eq_pulsetestset}
\end{equation}

\else
\mathindent=0mm
\begin{equation}
\begin{split}
SS_{t3} = &\bigl\{s_{t2}(f_{pulse}, w_{pulse}, f_{content}) \\
&\bigm| {f_{content}} \!=\! 440 , {f_{pulse}} \!=\! 4, w_{pulse} \!\in\! \{0.2, 0.5, 0.8\}\bigr\}
\label{eq_pulsetestset}
\end{split}
\end{equation}
\mathindent=7mm

\fi

%%% added 2021/05

\subsection{MFD-VL's Descriptiveness of Amplitude Envelopes of Simulated Environmental Sounds}
Environmental sounds such as chirping of insects and birds, and water streams have amplitude-modulated waveforms. It is well-known that some type of environmental sounds can be simulated by the granular synthesis technique~\cite{Keller1998a}. The granular synthesis technique splits a sound signal into small pieces called grains. Each grain has an envelope that contains an actual sonic content. If the content signal is a simple sinusoid, the synthesized sound signal is considered to be the same as a sound signal produced using the amplitude-modulation technique.

\begin{figure}[!t]
\centering
\includegraphics[width=3.3in]{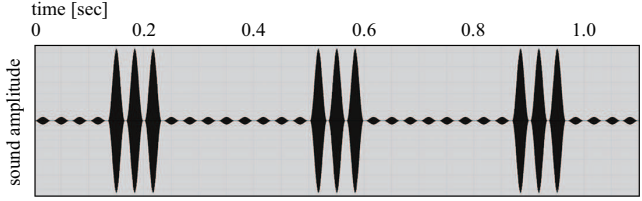}
\caption{Sound waveform of the signal $sim_{cricket2}(t, 5800, 3, 2.73, 30)$}
\label{fig_simcricket2waveform}
\end{figure}

For example, the sound signal of the chirping of a cricket has a carrier wave with a frequency of around 5~kHz, and the carrier wave is modulated into trains of syllable chirps whose rate is near 30~Hz~\cite{Thorson1982}. The sound of a cricket can be roughly simulated by the granular synthesis with an envelope created by the Hann window, which is defined as Eq.~(\ref{eq_hann}). Let $f_c$ be the frequency of the carrier wave and $f_e$ be the frequency of the envelope rate. The model of the chirping of a cricket $sim_{cricket}$ can be defined, as shown in Eq.~(\ref{eq_simcricketst}) and Eq.~(\ref{eq_simcricket}). The sound signal of the chirping of a {\it Gryllus bimaculatus}, which is a species of cricket living in Japan, can be simulated as $\mathop{sim_{cricket}\left(t, 5800, 30\right)}$. It is known that the chirping of a {\it Gryllus bimaculatus} sometimes has continuous 3-syllable chirps that are repeated three times per second. Let $f_c$ denote the frequency of the carrier wave, $sn$ be the number of continuous syllable chirps, $f_{rep}$ be the number of repetitions per second of the set of continuous syllable chirps, and $f_e$ be the frequency of continuous chirps. Then, the model of chirping of a cricket $sim_{cricket2}$ is defined using Eq.~(\ref{eq_MFD-VL-pulsecricketrect}) and Eq.~(\ref{eq_simcricket2}). Figure~\ref{fig_simcricket2waveform} shows the sound waveform of signal $sim_{cricket2}(t, 5800, 3, 2.73, 30)$.

\begin{equation}
\mathop{\omega\left(x\right) = \frac{1}{2}\left(1-\cos\left(2{\pi}x\right)\right)}
\label{eq_hann}
\end{equation}

\begin{equation}
\mathop{T_s\left(t, f\right)}=\min\left(t \bmod \frac{1}{f} , \frac{1}{1.1f}\right)
\label{eq_simcricketst}
\end{equation}

\begin{equation}
\mathop{sim_{cricket}\left(t, f_{c}, f_{e}\right)}=\sin\left(2{\pi}f_{c}t\right)\mathop{\omega\left(1.1f_{e}T_s\left(t, f_{e}\right)\right)}
\label{eq_simcricket}
\end{equation}

\ifhbonecolumn
\begin{equation}
rect_{cricket}({f_{e}},~sn,~f_{rep},~t)
= \begin{cases}
    1 & \left(t \bmod \frac{1}{f_{rep}} \leq \frac{sn}{f_{e}}\right) \\
    0.05 & (otherwise)
  \end{cases}
\label{eq_MFD-VL-pulsecricketrect}
\end{equation}

\begin{equation}
\mathop{sim_{cricket2}\left(t, f_{c}, sn, f_{rep}, f_{e}\right)} = rect_{cricket}({f_{e}},~sn,~f_{rep},~t) \mathop{sim_{cricket}\left(t, f_{c}, f_{e}\right)}
\label{eq_simcricket2}
\end{equation}

\else
\mathindent=0mm
\begin{equation}
rect_{cricket}({f_{e}},~sn,~f_{rep},~t) \!=\! \begin{cases}
    1 \!&\! \left(t \bmod \frac{1}{f_{rep}} \!\leq\! \frac{sn}{f_{e}}\right) \\
    0.05 \!&\! (otherwise)
  \end{cases}
\label{eq_MFD-VL-pulsecricketrect}
\end{equation}

\begin{multline}
\mathop{sim_{cricket2}\left(t, f_{c}, sn, f_{rep}, f_{e}\right)} \\
\!=\! rect_{cricket}({f_{e}},~sn,~f_{rep},~t) \mathop{sim_{cricket}\left(t, f_{c}, f_{e}\right)}
\label{eq_simcricket2}
\end{multline}
\mathindent=7mm
\fi

\begin{figure}[!t]
\centering
\includegraphics[width=3.3in]{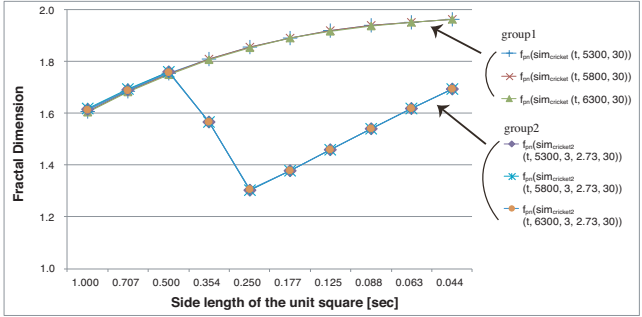}
\caption{Line charts of the MFD-VL signatures of simulated cricket sounds. These show the MFD-VL's robustness against fluctuations of a carrier signal's frequency.}
\label{fig_MFDVL-robust_freq}
\end{figure}

\begin{figure}[!t]
\centering
\includegraphics[width=3.3in]{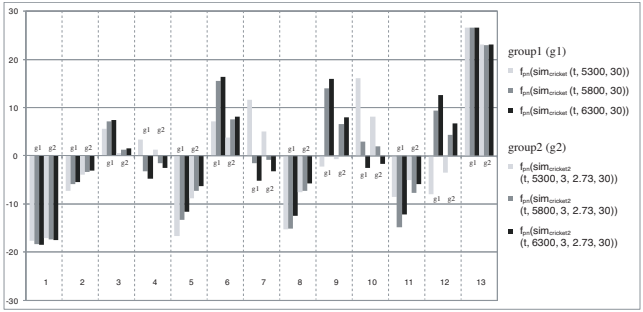}
\caption{MFCC13 values of simulated cricket sounds. These show the descriptiveness of varying carrier signals' frequency.}
\label{fig_MFCCs-robust_freq}
\end{figure}

We define the set of test sounds $SS_{t4}$ as Eq.~(\ref{eq_testset_sst4}). For preventing sparse spectra of the synthesized sound, each test sound is filtered by the pink noise filter function $f_{pn}$ Eq.~(\ref{eq_addpinknoise}), which is defined in Subsubsect.~\ref{subsubsec_MFD-VLdescbeatssw}. Figure~\ref{fig_MFDVL-robust_freq} shows the line charts of MFD-VL of $SS_{t4}$. Figure~\ref{fig_MFCCs-robust_freq} shows the 13-coefficients MFCCs values (MFCC13) of $SS_{t4}$. The SPTK toolkit~\cite{sptk} was used to compute MFCC13. The MFD-VL line charts of the sound set between $sim_{cricket}$ (group1) and $sim_{cricket2}$ (group2) are clearly different. These results indicate that the MFD-VL signature can clearly discriminate the two amplitude envelope signals.

\ifhbonecolumn
\begin{multline}
SS_{t4} = \left\{f_{pn}\bigl(sim_{cricket}(t, f, 30)\bigr) \bigm| f \in \{5300, 5800, 6300\}\right\} \\
\cup \left\{f_{pn}\bigl(sim_{cricket2}(t, f, 3, 2.73, 30)\bigr) \bigm| f \in \{5300, 5800, 6300\}\right\}
\label{eq_testset_sst4}
\end{multline}

\else
\mathindent=0mm
\begin{equation}
\begin{split}
SS_{t4} \!=\! &\bigl\{f_{pn}\bigl(sim_{cricket}(t, f, 30)\bigr) \bigm| f \!\in\! \{5300, 5800, 6300\}\bigr\} \\
&\cup \bigl\{f_{pn}\bigl(sim_{cricket2}(t, f, 3, 2.73, 30)\bigr) \\
&\quad \bigm| f \!\in\! \{5300, 5800, 6300\}\bigr\}
\label{eq_testset_sst4}
\end{split}
\end{equation}
\mathindent=7mm

\fi

\subsection{MFD-VL's Robustness against Fluctuations of a Carrier Signal's Frequency}
We have assumed that one of the main causes of the diversity of environmental sounds is small fluctuations of sound source parameters, such as carrier signal frequency, due to the individuality of the sound source (D1). For example, the frequency of the human voice varies with the size and form of the individual's vocal tract which are the sound source parameters with fluctuations. However, we can recognize the sound recordings of talking by children, adults, and aged persons as the human voice in the same manner. The set of test sounds $SS_{t4}$ contains the synthesized sound signals of the chirping of a cricket with different frequencies of a carrier wave. We can evaluate the descriptiveness of acoustic features for the small fluctuations of sound source parameters by comparing the feature vectors of test sounds contained in $SS_{t4}$.

Figure~\ref{fig_MFDVL-robust_freq} shows the robustness of MFD-VL against the fluctuation of the carrier signal's frequency contained in the amplitude envelopes. We defined the discrimination rate of both MFD-VL and MFCC13 for the quantitative evaluation. Let $se(ts, n)$ denote the \mbox{$n$-th} element value of MFD-VL signature of the test sound clip $ts$. The MFD-VL discrimination rate $DR_{mfdvl}$ for comparing two test sounds A and B is calculated using Eq.~(\ref{eq_disratemfdvl}). The $RANGE_{mfdvl}$ in Eq.~(\ref{eq_disratemfdvl}) denotes the width of the range of $se(ts, x)$, which may take a value between 1 and 2. Therefore, $RANGE_{mfdvl}$ is calculated as 1 $(=2-1)$. The MFCC13 discrimination rate $DR_{mfcc13}$ for comparing two test sounds A and B was defined as Eq.~(\ref{eq_disratemfcc13}) where $se2(ts, n)$ was defined as the \mbox{$n$-th} dimension value of the MFCC13 vector of the test sound clip $ts$. The constant $RANGE_{mfcc13}$ was set to 56.3. This number was determined from the difference between the minimum and the maximum values of MFCC13 of 3000 environmental sounds in a dataset for the experimental evaluation which is described in Sect.~\ref{sec_exp_eval}. The sounds in the dataset were collected from the Freesound project~\cite{freesound}.

\begin{equation}
DR_{mfdvl}=\sqrt{\frac{1}{10} \sum_{n=1}^{10}\left(\frac{se(A,n) - se(B,n)}{RANGE_{mfdvl}}\right)^2}
\label{eq_disratemfdvl}
\end{equation}

\begin{equation}
\mathop{DR_{mfcc13}}=\sqrt{\frac{1}{13} \sum_{n=1}^{13}\left(\frac{se2(A,n) - se2(B,n)}{RANGE_{mfcc13}}\right)^2}
\label{eq_disratemfcc13}
\end{equation}

Table~\ref{fig_discrimination_rates_robust_freq} lists the comparisons of the two discrimination rates $DR_{mfdvl}$ and $DR_{mfcc}$ to distinguish two test sounds with various carrier signal frequencies which are contained in $SS_{t4}$. This comparison shows that to distinguish any two test sounds with various carrier signal frequencies, the values of $DR_{mfdvl}$ are not more than 0.22\%, and those of $DR_{mfcc}$ are not less than 2.90\% and no more than 16.15\%. These results indicate that the MFD-VL signature shows robustness against the fluctuation of the carrier signal frequency, which is the diversity cause D1 defined in Subsect.~\ref{subsec_acousticfeatureextraction_for_esr}. The MFD-VL signature can clearly discriminate the characteristics of the amplitude envelopes with its robustness against the fluctuation of the carrier signal frequency.

\begin{table}[!t]
\centering
\caption{Comparison of discrimination rates of varying carrier signals' frequencies between the MFD-VL signature and MFCC13}
\label{fig_discrimination_rates_robust_freq}
\includegraphics[width=3.3in]{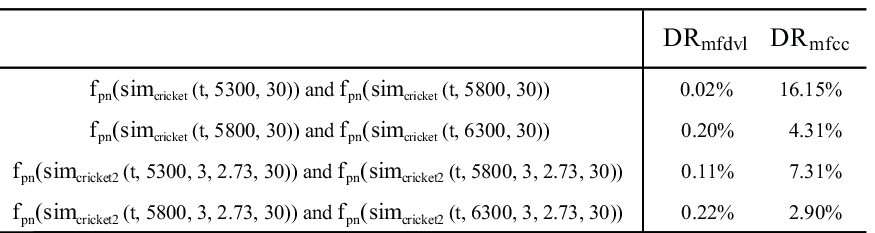}
\end{table}

\subsection{MFD-VL's Robustness against Noises}
To evaluate the robustness of MFD-VL against noises, we use test sounds generated from the simulated cricket sound and pink noise signals. Let $S_{cricket2}$ denote $sim_{cricket2}(t,~5800,~3,~2.73,~30)$, $Noise_{pink}$ denote a pink noise signal whose maximum amplitude is normalized to -0.1~db, and $\beta$ be a constant number to control the signal-to-noise ratio (SNR). Then, a set of test sounds $SS_{t5}$ is calculated using Eq.~(\ref{eq_noisetestset}). The set of constant numbers $\beta\in\{0.67, 0.8, 0.89, 0.94, 0.97\}$ corresponds to the set of SNRs $\{6db, 12db, 18db, 24db, 30db\}$ of $SS_{t5}$. The frequency components below 40 Hz contained in the pink noise are cut off by using a low cut filter before the amplitude normalization because the components with lower frequencies cannot be recorded nor played using common microphones and speakers.

\ifhbonecolumn
\begin{equation}
SS_{t5} = \bigl\{\beta~S_{cricket2} + (1 - \beta)~Noise_{pink} \bigm| \beta\in\{0.67, 0.8, 0.89, 0.94, 0.97\}\bigr\}
\label{eq_noisetestset}
\end{equation}

\else
\begin{multline}
SS_{t5} = \bigl\{\beta~S_{cricket2} + (1 - \beta)~Noise_{pink} \\
\bigm| \beta\in\{0.67, 0.8, 0.89, 0.94, 0.97\}\bigr\}
\label{eq_noisetestset}
\end{multline}

\fi

\begin{figure}[!t]
\centering
\includegraphics[width=3.3in]{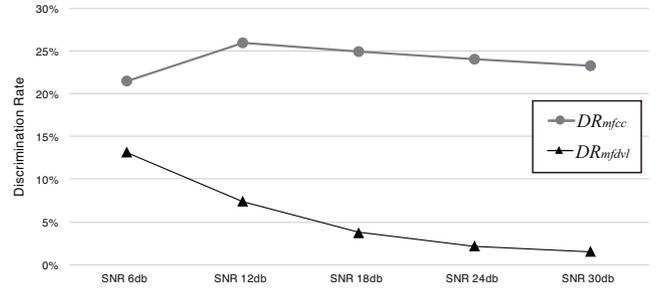}
\caption{Line charts of the discrimination rates $DR_{mfdvl}$ and $DR_{mfcc}$ to distinguish the simulated cricket sound $S_{cricket2}$ and each sound in the set of test sounds $SS_{t5}$.}
\label{fig_robust_noise}
\end{figure}

Figure~\ref{fig_robust_noise} shows the line charts of the discrimination rates $DR_{mfdvl}$ and $DR_{mfcc}$ for distinguishing the simulated cricket sound $S_{cricket2}$ and each sound in the set of test sounds $SS_{t5}$. The results indicate that the MFD-VL signature has the higher robustness against the pink noise than MFCC13, and that the MFD-VL signature is not relatively affected by the pink noise. It is confirmed that the MFD-VL signature is certainly robust against the diversity cause D2 defined in Subsect.~\ref{subsec_acousticfeatureextraction_for_esr}.

%%% added

\section{Experimental Evaluation (EXP1)}\label{sec_exp_eval}
\subsection{Experimental Setups}\label{subsec_expsetup}
To evaluate a descriptiveness of the acoustic feature signatures based on the multiscale fractal dimension, we developed a similarity search system using the k-nearest neighbors (k-NN) method. We examined the performances of the similarity search tasks by using the proposed acoustic feature signatures.

To produce a sound dataset of environmental sounds, sound samples were collected from the Freesound project, which has stored many types of environmental sounds uploaded by users worldwide. The Freesound allows users to share their sounds and describe metadata regarding their shared sounds on the web. Each sound is labeled with a group of tags that are relatively well maintained as user generated contents~\cite{Akkermans2011}. The tags represent the objects the users were listening to in their listening experiences. Based on the following rules defined by the authors, the sounds were chosen and imported to our dataset to be used in the similarity search system. The sounds that were tagged with ``field-recording'' and with lengths between 1 and 600~s were chosen. This means that these sound clips have been recorded outside a recording studio. Each sound source outside the studio is typically unique and never the same. Therefore, sound clips that are tagged with "field-recording" have the diversity cause D1.  We imported the top 3000 sounds in the descending order of downloaded number by unspecified users counted by the Freesound's system for each sound. Each sound was converted to a uniform format (1 channel, 44100~Hz sampling frequency, 16~bits bit depth, and maximal amplitude normalized to -0.1~db) for normalization before extracting acoustic features including EMFD, EMFD-KDE, MFD-VL, and MFCCs. The average length of the imported sounds is 70.4~s.

\subsection{Acoustic Feature Extraction}
One of the most well-known acoustic features used for ESR is MFCCs. Here, we used MFCCs for comparing the descriptiveness with our newly-proposed feature signatures. The SPTK toolkit was used to compute 13-coefficients MFCCs (MFCC13) and MFCC39, which represents the first and the second-order derivatives of MFCC13. MFCC13 and MFCC39 were computed using a fixed width analysis window of length 50~ms. The feature sets of MFCC13 and MFCC39 consist of mean values of their coefficients of the analysis window. EMFD and EMFD-KDE consist of the 512 elements defined in Sect.~\ref{sec_EMFD} and \ref{sec_EMFD-KDE}, and MFD-VL consists of the 10 elements defined in Sect.~\ref{sec_MFD-VL}. In this experiment (EXP1), we determined that the smoothing parameter $\alpha$ used for computing the EMFD-KDE signature in Eq.~(\ref{eq_bwEMFD}) is 32.

%%%%%% Through the experiments with different values of $\alpha$, we found that the best result for the similarity search is obtained for $\alpha = 32$. In this study, $h_{rbin}(32)$ is used as the bandwidth for each radius of the unit disk to compute the EMFD-KDE signature.

Table~\ref{fig_features} lists different feature sets to be compared through experimental evaluation. L1 represents the total number of features in the concatenated feature sets and L2 represents the number of features in the feature sets after dimensionality reduction through principal components analysis (PCA). To achieve the best possible performance of the similarity search using k-NN method, PCA was applied to the feature vectors of the most frequently downloaded 600 sounds in the dataset to extract its eigenvectors for dimensionality reduction. The ``prcomp'' function of R language was used for PCA processing. The corresponding L2s of feature sets 1, 3, 6, and 8 were determined so that each of their cumulative contribution ratios was 99\%. The L2s of feature sets 2, 4, 5, 7, 9, and 10 were fixed to 114.

\begin{table}[!t]
\centering
\caption{List of acoustic feature sets for the comparison of their descriptiveness.}
\label{fig_features}
\includegraphics[width=3.3in]{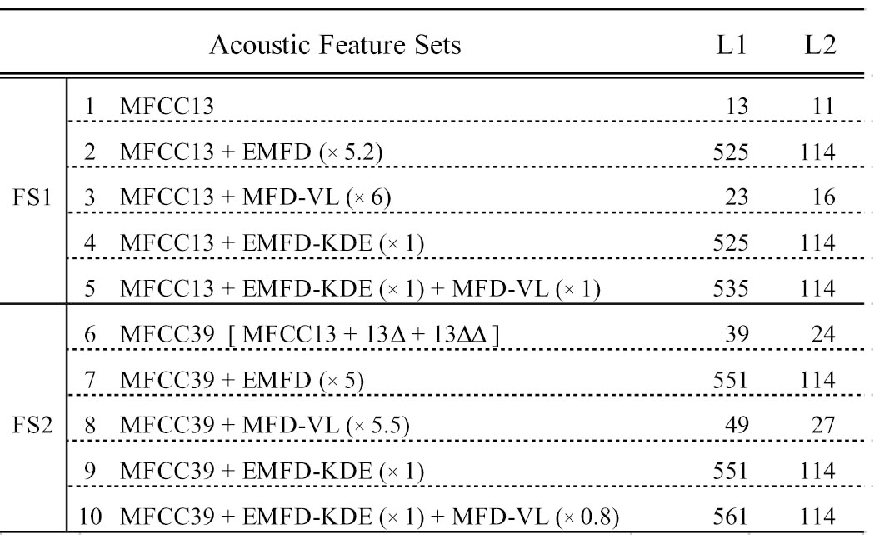}
\end{table}

Feature sets 1 and 6 are the standard sets for comparing with other feature sets. Each feature set from 2 to 5 consists of MFCC13, and each feature signature is based on the multiscale fractal dimension. We defined that each feature set from 1 to 5 belongs to the group FS1. In addition, each feature set from 7 to 10 consists of MFCC39 and each feature signature is based on the multiscale fractal dimension. Moreover, we defined each feature set from 6 to 10 to belong to group FS2.

The suffix ``($\times~\gamma$)'' of each feature vector denotes a weighting coefficient $\gamma$. Each value of the feature vectors is multiplied by $\gamma$ when its feature vector is combined with other feature(s). Through the experimental evaluation, the weighting coefficient $\gamma$ for each feature vector was appropriately chosen to perform the best result.

\subsection{Evaluation Method}
The similarity search system using k-NN method returns a search result list of environmental sounds based on the distance in the space of the selected feature set through a search-key sound. To evaluate the performance using each feature vector in table~\ref{fig_features}, we defined the similarity index $SI$ between the tag group of the search-key sound $tags_{key}$ and that of the retrieved sound $tags_{s}$ as Eq.~(\ref{eq_simidx}). This index is known as the Jaccard similarity coefficient that measures similarity between finite sample sets.

\begin{equation}
SI = \frac{\mathrm{card}\left(tags_{key} \cap tags_{s}\right)}{\mathrm{card}\left(tags_{key} \cup tags_{s}\right)}
\label{eq_simidx}
\end{equation}

To improve an accuracy of similarity index $SI$, we removed the commoner morphological and inflexional endings from all tags by using Porter Stemmer~\cite{Porter1980} in advance. Furthermore, the predefined stop words include sound formats, such as ``mp3'' and ``stereo,'' and tool makers, such as ``sony'' and ``tascam,'' were removed from the tag groups for computing the $SI$s. The tags that contain the text ``fieldrecord'' were removed from the tag groups because all sounds in the dataset have them. For each of the 3000 sounds in the dataset, $SI$s between a search-key sound and each retrieved sound in the search-result list were computed. We compared the average values of the $SI$s of 3000 sounds for each acoustic feature set.

\subsection{Evaluation Results}
Table~\ref{fig_evalres} shows the $SI$s of ``top $n$'' for each feature set. The $SI$ of ``top $n$'' is the average value of the $SI$s between a search-key sound and each retrieved sound in the top $n$ rank of the search-result list. Table~\ref{fig_evalres} shows the average values of the $SI$s of ``top $n$'' computed for each of the 3000 sounds. For reference, the average value of the $SI$ between two randomly chosen sounds in the dataset is 0.014.

\begin{table}[!t]
\centering
\caption{Evaluation results of each feature set quantified using $SI$s}
\label{fig_evalres}
\includegraphics[width=3in]{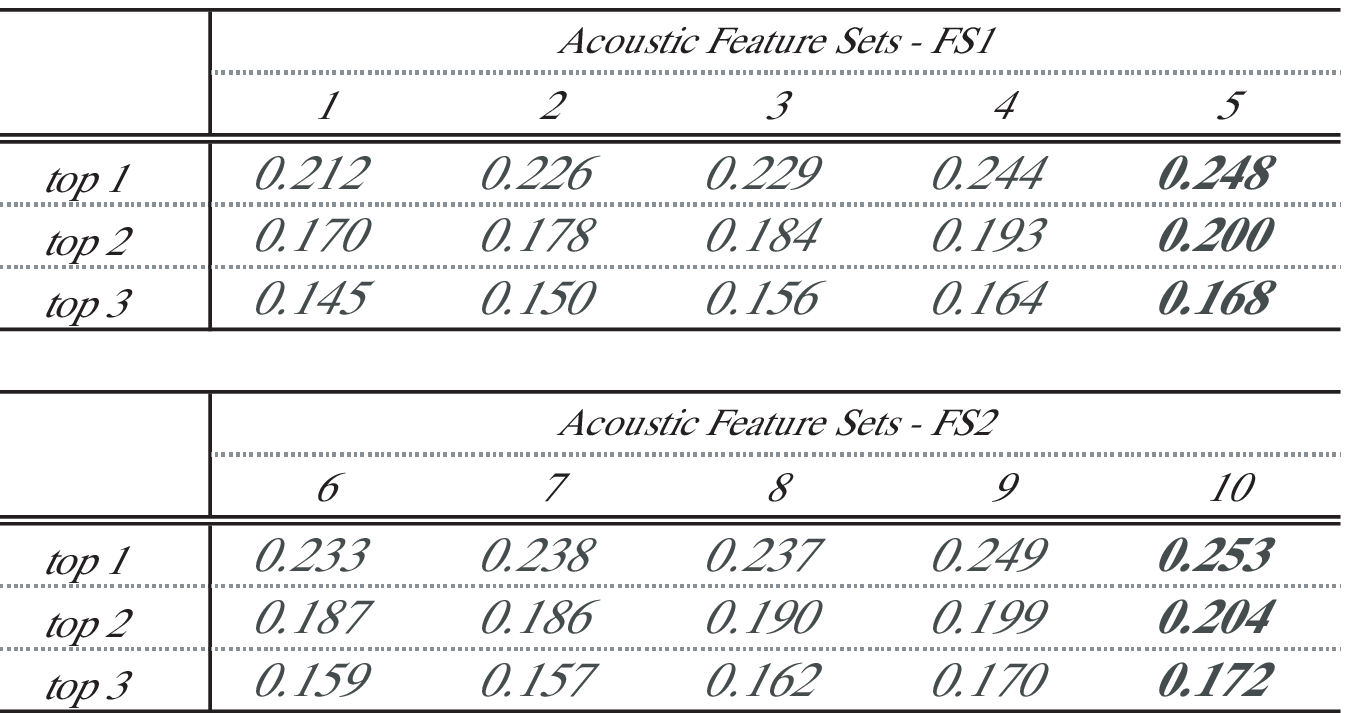}
\end{table}

%\ifhbonecolumn %%%%%
%\begin{figure*}[!t]
%\centering
%\includegraphics[width=0.88\textwidth]{sunou18}
%\caption{Evaluation results of each feature set quantified using $SI$s}
%\label{fig_evalres}
%\end{figure*}
%
%\else %%%%%
%\begin{figure*}[!t]
%\centering
%\includegraphics[width=0.88\textwidth]{sunou18}
%\caption{Evaluation results of each feature set quantified using $SI$s}
%\label{fig_evalres}
%\end{figure*}
%
%\fi %%%%%

By comparing feature sets 4 and 9 with 2 and 7, respectively, it was confirmed that the descriptiveness of EMFD-KDE is superior to that of EMFD. $SI$ of ``top 1'' of feature set~4 was 8.1\% higher than that of feature set~2. $SI$ of ``top 1'' of the feature set~9 was 4.9\% higher than that of feature set~7.

By comparing feature sets~3 and 8 with 1 and 6, respectively, it was confirmed that the newly-developed MFD-VL signature improves the performance of similarity search result. The MFD-VL signature can describe some acoustic features that MFCCs cannot. $SI$ of ``top 1'' of the feature set~3 was 8.2\% higher than that of the feature set~1. $SI$ of ``top 1'' of feature set~8 was 1.6\% higher than that of the feature set~6.

The results obtained using the feature sets~5 and 10 achieved the best similarity search performance in each group of feature vectors FS1 and FS2. $SI$ of ``top 1'' of feature set~5 was 17.2\% higher than that of the feature set~1. $SI$ of ``top 1'' of the feature set~10 was 8.7\% higher than that of the feature set~6. It was confirmed that EMFD-KDE and MFD-VL are effective as an acoustic feature signature for the environmental sounds.

\section{Experimental Evaluation Using Public Dataset and Other Acoustic Features (EXP2)}\label{sec_expevaldcase2018}
\subsection{Experimental Setups}\label{subsec_expsetup2}
We developed another similarity search system to evaluate the proposed acoustic feature signatures by comparing them with the top-ranked acoustic features used in Task2 of the DCASE 2018 challenge~\cite{Fonseca2018}. As a sound dataset of this similarity search system, we used the train set of "Freesound Dataset Kaggle 2018" (FSDKaggle2018) ~\cite{Fonseca2019} which contains 9473 sound clips provided as uncompressed PCM 16 bit, 44.1 kHz, mono audio files. The significant characteristics of the dataset are as follows:
\begin{itemize}
\item The sound clips are unequally distributed in the 41 categories of Google's AudioSet Ontology ~\cite{Gemmeke2017}. The minimum number of sound clips per category is 94, and the maximum is 300.
\item Each sound clip is annotated with a single ground-truth label of the categories.
\item The duration of the sound clips ranges from 300ms to 30s.
\end{itemize}
The maximal amplitude of sound clips was normalized to -0.1db before the acoustic feature extraction.

\subsection{Acoustic Feature Extraction}
We chose the top-ranked acoustic features proposed in the DCASE 2018 challenge task2, including log-mel energies, Perceptual weighted power spectrogram, and Logarithmic-filtered log-spectrogram, to compare their descriptiveness with that of the EMFD-KDE and MFD-VL signatures. Table~\ref{fig_features_e2} shows the acoustic feature sets to be used in EXP2. The significant processes of the feature extraction are as follows:
\begin{itemize}
\item All acoustic features except MFD-VL were computed using fixed-width analysis windows of length 50ms every 25ms. 
\item For computing the MFD-VL signature, the length of the sound signal must be longer than 1s. In case the target sound clip is shorter than or equal to 1s, we repeated the sound signal so that the total length of the repeated signals becomes longer than 1s.
\item Let the smoothing parameter $\alpha$ be 1, which is used for computing the EMFD-KDE signature in Eq.~(\ref{eq_bwEMFD}).
\item Dimensionality reduction was performed through PCA using \textit{scikit-learn} library. The numbers of features L2s after dimensionality reduction were determined so that each of their cumulative contribution ratios was 98\%.
\end{itemize}

\begin{table}[!t]
\centering
\caption{List of acoustic feature sets for the comparison of their descriptiveness.}
\label{fig_features_e2}
\includegraphics[width=3.3in]{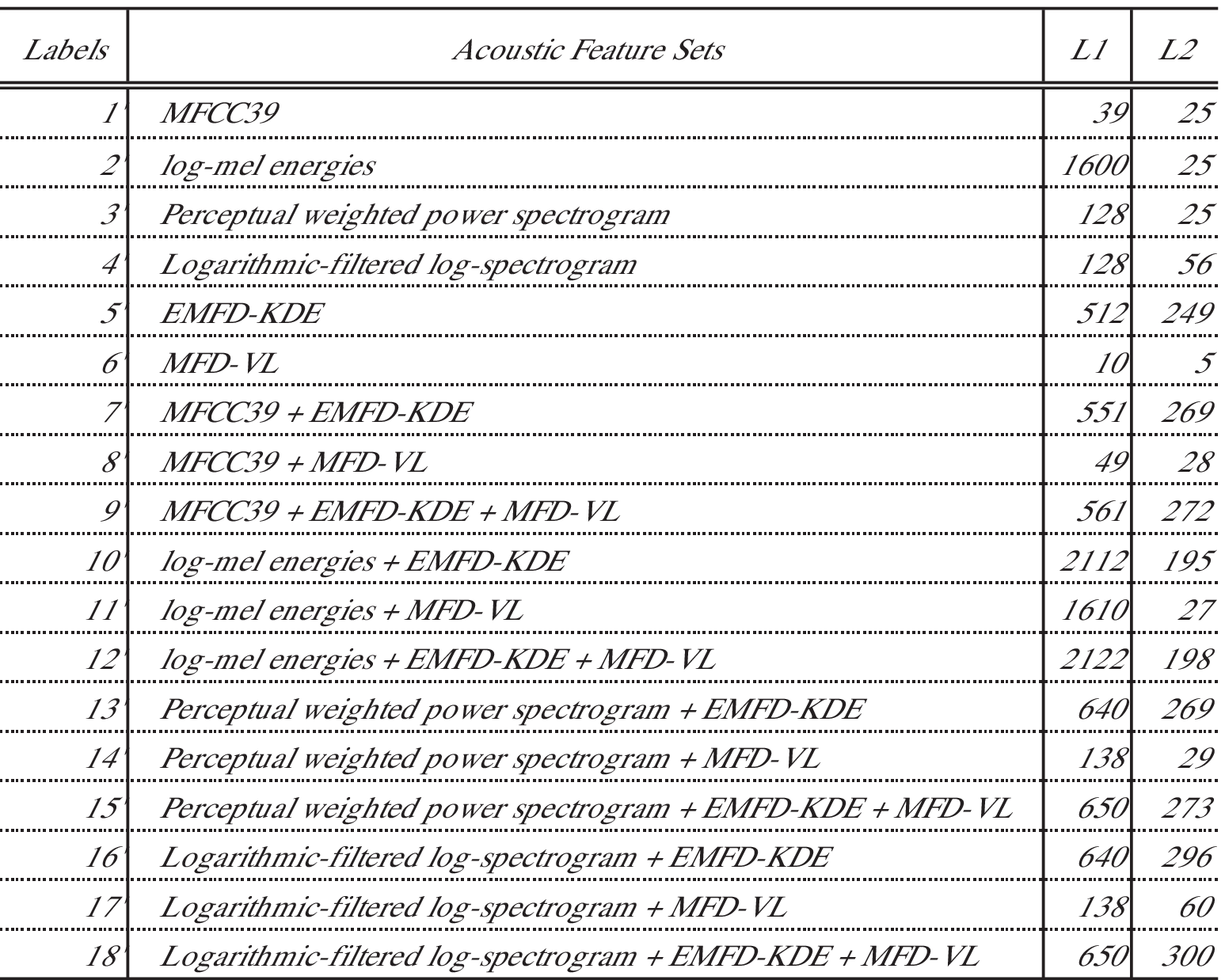}
\end{table}

\subsection{Evaluation Method}
For each of the 9473 sounds in the dataset, the relevances between the search-key sound and each retrieved sound in the search-result list were computed using their category labels. Then, we computed the mean value of Precision@k for each query on the entire dataset to measure the top-$k$ ranking quality of the search result lists.

\subsection{Evaluation Results}
Table~\ref{fig_mP} shows the Precision@k scores at ranking positions 1, 3, and 10 on the entire dataset. Bold numbers are the best Precision@k scores for each ranking position $k$. The results obtained using the feature set 8' (MFCC39 + MFD-VL) achieved the best similarity search performance.
By comparing feature sets (2' with 10', 11', and 12'), (3' with 13', 14', and 15'), and (4' with 16', 17', and 18') respectively, it was confirmed that both the EMFD-KDE and MFD-VL signatures improve the performance of the similarity search task.
By comparing feature sets 1' to 6', we confirmed that the Precision@10 score obtained using the feature set 5' (EMFD-KDE) achieved better performance than those obtained using the feature sets 2' (log-mel energies), 3' (Perceptual weighted power spectrogram), and 4' (Logarithmic-filtered log-spectrogram), respectively.

\begin{table}[!t]
\centering
\caption{Precision@k scores at ranking positions 1,3, and 10 on the entire dataset.}
\label{fig_mP}
\includegraphics[width=3.3in]{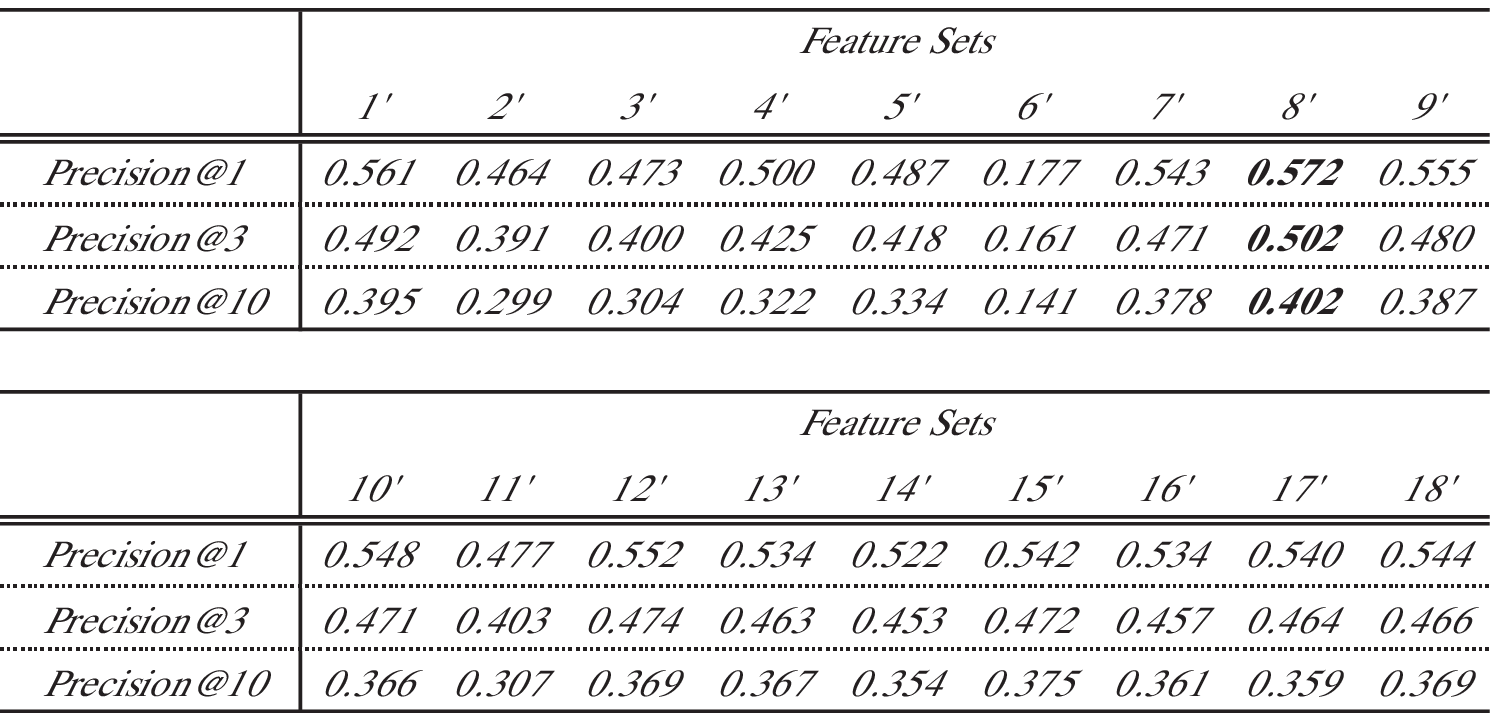}
\end{table}

Table~\ref{fig_mP3_categ} shows the Precision@3 scores of the search-result list grouped by the 41 categories for each feature set. Bold numbers are the best Precision@3 scores for each category. The top five numbers of the categories in which each feature set obtained the best Precision@3 score are 10 categories for feature set 8' (MFCC39 + MFD-VL), 9 categories for feature set 1' (MFCC39), 7 categories for feature set 9' (MFCC39 + EMFD-KDE + MFD-VL), 5 categories for feature set 17' (Logarithmic-filtered log-spectrogram + MFD-VL), and 4 categories for feature set 18' (Logarithmic-filtered log-spectrogram + EMFD-KDE + MFD-VL).

\begin{table*}[!t]
\centering
\caption{Precision@3 scores for each category.}
\label{fig_mP3_categ}
\includegraphics[width=0.95\textwidth]{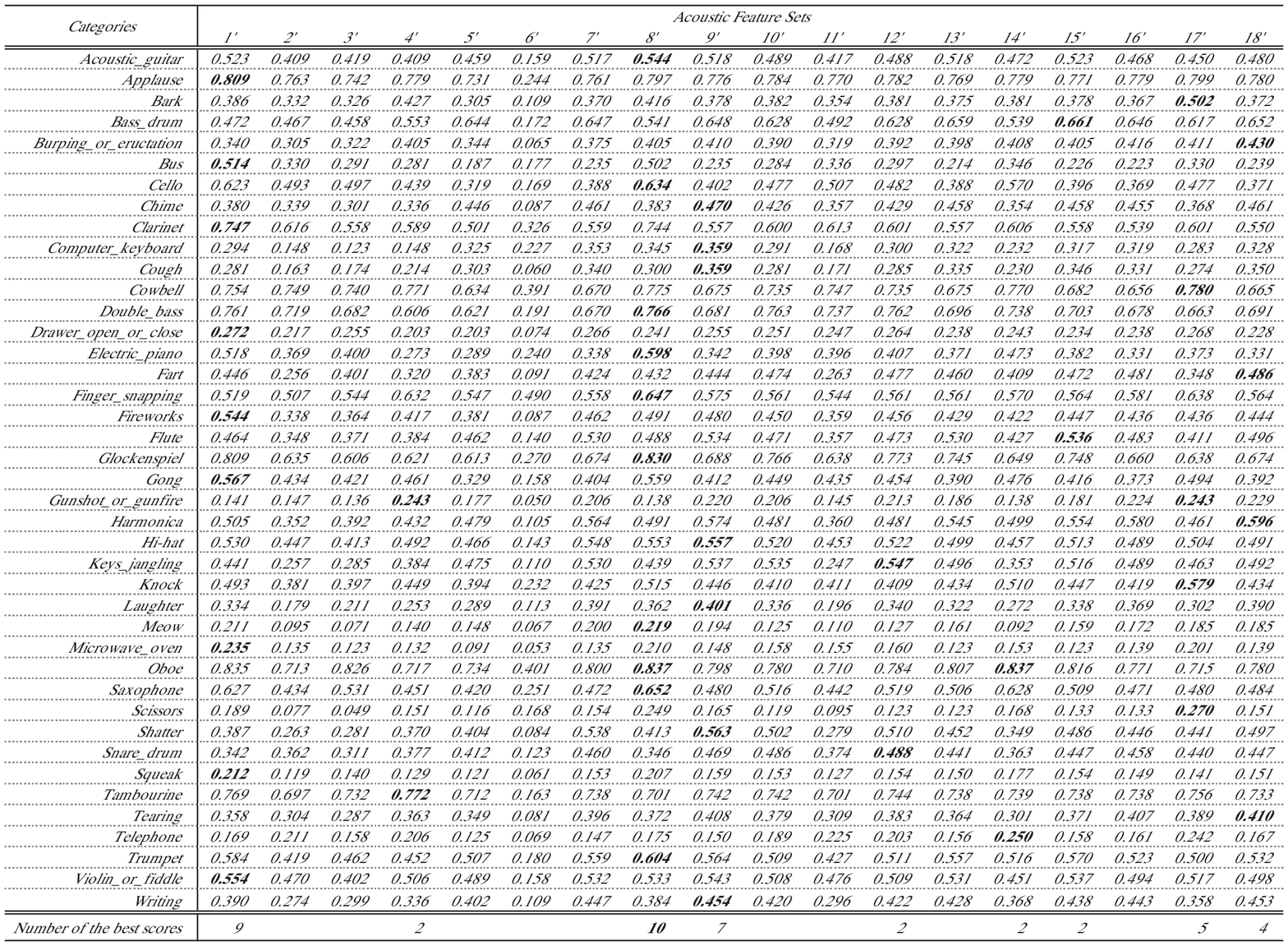}
\end{table*}

\section{Conclusion}\label{sec_conclusion}

Recent research on ESR focused on the evaluation of time-domain features of environmental sounds. For ESR, an acoustic feature must describe the nonstationary characteristics of target sounds as time-domain features and must be robust against the following three main causes of the diversity of environmental sounds.
\begin{itemize}
\item[D1)]Small fluctuations of sound source parameters, such as carrier signal frequency, due to the individuality of the sound source.
\item[D2)]Background noises that the person who recorded the target sound did not expect to record.
\item[D3)]Mixed composition of different types of sound sources.
\end{itemize}
In this study, we have focused on the extraction of acoustic feature signatures that are robust against the diversities caused by D1 and D2.

In a previous study, we proposed the EMFD feature signature to describe both frequency- and time-domain features of target sounds.

However, we also recognized the following problems in EMFD.
\begin{itemize}
\item[p1)]EMFD includes error values.
\item[p2)]It is oversensitive to discriminate the features of environmental sounds.
\item[p3)]It lacks the robustness against the diversity of environmental sounds.
\item[p4)]It cannot describe the time-domain features for periods longer than 10~ms.
\end{itemize}
To solve these problems, we proposed the EMFD-KDE and MFD-VL feature signatures.

The newly-proposed EMFD-KDE feature signature is the probability distribution of the enhanced MFD values at each radius of the unit disk computed using the kernel density estimation method. In Subsect.~\ref{subsec_opt_bandwidth}, we studied the method to optimize bandwidth $h$ used for kernel density estimation. Based on the normal reference rule, we defined bandwidth $h_{rbin}(\alpha)$ optimized for each radius of the unit disk by using Eq.~(\ref{eq_bwEMFD}).
Through the experiments with different values of the smoothing parameter $\alpha$, we determined that the best result of the similarity search task EXP1 is obtained for $\alpha = 32$ and that of EXP2 is obtained for $\alpha = 1$. We assume that this difference of the optimized value of $\alpha$ for each task is caused by the diversity difference of datasets for each task. In the FSDKaggle2018 dataset for EXP2, a number of the ground truth labels have been manually verified, the number of categories is only 41, and the length of the sound clip is shorter than 30sec. The dataset for EXP1 is more diverse and complex than the FSDKaggle2018 dataset.

In Sect.~\ref{sec_MFD-VL}, we proposed the MFD-VL signature and demonstrated its characteristics through experiments using the simulated cricket's sounds as follows.
\begin{itemize}
\item MFD-VL can discriminate the frequencies of the amplitude envelopes between 22.6 and 1 Hz.
\item It can discriminate the shapes of the amplitude envelopes.

\item It is robust against the fluctuation of the carrier signal frequency (, i.e., the diversity cause D1 defined in Subsect.~\ref{subsec_acousticfeatureextraction_for_esr}).
\item It is robust against background noises (, i.e., the diversity cause D2 defined in Subsect.~\ref{subsec_acousticfeatureextraction_for_esr}).

\end{itemize}
The MFD-VL signature is expected to describe the time-domain features for periods longer than 10~ms. The MFD-VL signature shows stability and robustness against background noises and small fluctuations of the carrier signal's frequency.

From the experimental evaluation results (EXP1), we confirmed that the descriptiveness of EMFD-KDE supplementing MFCC13 and MFCC39 is evidently higher than that of EMFD supplementing MFCC13 and MFCC39. We conclude that the smoothness of the EMFD-KDE signature can solve problems p1, p2, and p3. Furthermore, the experimental evaluation results showed that the MFD-VL signature supplementing EMFD-KDE and MFCCs improves the performance of the similarity search. The MFD-VL signature functions as an effective time-domain feature and can solve problems p3 and p4.

In Sect.~\ref{sec_expevaldcase2018}, we conducted another experimental evaluation (EXP2) using the FSDKaggle2018 dataset and other acoustic features. We confirmed clear evidence that both the EMFD-KDE and MFD-VL signatures have the unique descriptiveness of environmental sound. These signatures are effective when they are used with other acoustic features, including MFCC39 and the top-ranked acoustic features in the DCASE 2018 challenge task2.

The EMFD-KDE signature has 512 feature elements, which is  relatively  more than those of other feature signatures. This implies that the EMFD-KDE signature requires more computational time than the conventional methods for feature extraction. For the similarity search task, feature signatures of library sounds can be computed \textit{a priori} which implies that we do not care about the computation time of the EMFD-KDE signature. While the EMFD-KDE also requires more searching time than conventional method, this time is negligible compared to the time required for the retrieval of matched sounds from the library. The computational time of EMFD-KDE and the length of searching time using EMFD-KDE do not matter.

Environmental sounds have the acoustic features in their frequency domain, as well as other important features in the time domain with various time scales. We conclude that both the EMFD-KDE and MFD-VL signatures can describe the essential acoustic features of environmental sounds with robustness against the diversity of environmental sounds. Further studies
are needed to evaluate the performance of other applications using these acoustic feature signatures, such as classification tasks using machine-learning systems.

%%%%%%%%%%%%%%%%%%%%%%%%%%%%%%%%%%%%%%%%

\section*{Acknowledgments}
This work was supported by JSPS KAKENHI Grant Number 18K11378.

%\bibliographystyle{ieicetr}% bib style
%\bibliography{}% your bib database

%\profile{}{}
%\profile*{}{}% without picture of author's face

\profile[sunou]{Motohiro Sunouchi}{received masters of environmental studies in human and engineered environmental studies from the University of Tokyo in 2004. He is currently pursuing the Ph.D. in information science at the Hokkaido University. He has been a research assistant since 2007 and a senior lecturer since 2016 with the Department of Design, Sapporo City University, Japan. His research interests lie in the areas of audio signal processing and auditory culture.}
\profile[yoshioka]{Masaharu Yoshioka}{Masaharu Yoshioka, Ph. D is a Professor of Faculty of Information Science and Technology, Global Station of Big Data and Cybersecurity, and Institute for Chemical Reaction Design and Discovery of Hokkaido University. He received the B.E. and M.E. degrees of precision engineering and the Ph.D. degree of precision machinery engineering from University of Tokyo, Japan, in 1991, 1993, and 1996, respectively. From April 1996 to March 2000, he was a Research Associate of National Center for Science and Information Systems, Japan. From April 2000 to May 2001, he was a Research Associate of National Institute of Informatics, Japan. From June 2001, he joined the Graduate School of Engineering as a Associate Professor and this school is reorganized as Graduate School of Information Science and Technology in 2004. From January 2019, he became a Professor of Faculty of Information Science and Technology. He also joined Institute for Chemical Reaction Design and Discovery from January 2020. From October 2017, he also serve as a visiting researcher at RIKEN Center for Advanced Intelligence Project. His research interests includes application of knowledge engineering technology for information access and knowledge management, Linked Open Data, and application of knowledge engineering technology for a particular research domain (e.g., cheminformatics and nanoinformatics).}

\end{document}